\documentclass[
preprint,
preprintnumbers,
nofootinbib,
prd,aps
]{revtex4}

\usepackage{amssymb}
\usepackage{amsmath}
\usepackage{subfigure}
\usepackage{graphicx}
\usepackage{hyperref}
\usepackage{bm}
\usepackage{braket}

\begin{document}

\newcommand{\rem}[1]{$\spadesuit${\bf #1}$\spadesuit$}
\newcommand{\red}[1]{{\textcolor{red}{#1}}}
\newcommand{\blue}[1]{{\textcolor{blue}{#1}}}
\newcommand{\figref}[1]{Fig.\ref{#1}}
\newcommand{\tabref}[1]{Table\ref{#1}}

\preprint{OU-HET 1117}

\title{
Strongly first-order electroweak phase transition\\
by relatively heavy additional Higgs bosons
}

\author{
Shinya Kanemura\footnote{kanemu@het.phys.sci.osaka-u.ac.jp}
and
Masanori Tanaka\footnote{m-tanaka@het.phys.sci.osaka-u.ac.jp}
}

\affiliation{
Department of Physics, Osaka University, Toyonaka, Osaka 560-0043, Japan
}

\begin{abstract}

We discuss first-order electroweak phase transition in models with extended Higgs sectors for the case with relatively heavy additional scalar bosons. We first show that, by the combination of the sphaleron decoupling condition, perturbative unitarity and vacuum stability, 
mass upper bounds on additional scalar bosons can be obtained at the TeV scale even at the alignment limit where the lightest Higgs boson behaves exactly like the SM Higgs boson at tree level. We then discuss phenomenological impacts of the case with the additional scalar bosons with the mass near 1~TeV. Even when they are too heavy to be directly detected at current and future experiments at hadron colliders, the large deviation in the triple Higgs boson coupling can be a signature for first-order phase transition due to quantum effects of such heavy additional Higgs bosons. On the other hand, gravitational waves from the first-order phase transition are found to be weaker in this case as compared to that with lower masses of additional scalar bosons.

\end{abstract}

\maketitle
\renewcommand{\thefootnote}{\arabic{footnote}}

\section{Introduction}

Although the standard model (SM) has been successful being consistent with current data at LHC~\cite{ATLAS:2019nkf, CMS:2020gsy}, there are phenomena that cannot be explained in the SM such as neutrino oscillation, dark matter and baryon asymmetry of the Universe (BAU). Therefore, new physics beyond the SM is absolutely necessary.

In order to explain the BAU, the idea of baryogenesis is the most promising. A new model is required to satisfy the Sakharov's conditions to realize baryogenesis~\cite{Sakharov:1967dj}. It has turned out that the SM cannot satisfy these conditions~\cite{Kajantie:1996mn, Kajantie:1996qd, DOnofrio:2015gop, Huet:1994jb}. Its extension has to be considered for successful baryogenesis. In particular, for the scenario of electroweak baryogenesis~\cite{Kuzmin:1985mm}, extended Higgs sectors are often introduced to satisfy the Sakharov's conditions having the sufficient amount of CP violation and realizing strongly first-order electroweak phase transition.

The strongly first-order phase transition is the most important characteristic property for the model of electroweak baryogenesis. The electroweak phase transition in extended Higgs models with additional doublet scalar fields~\cite{Turok:1991uc, Cottingham:1995cj, Cline:1996mga, Kanemura:2004ch, Fromme:2006cm, Moreno:1996zm, Funakubo:2009eg, Basler:2016obg, Bernon:2017jgv, Dorsch:2017nza, Barman:2019oda}, singlet fields with scalar mixing~\cite{Ahriche:2007jp, Fuyuto:2014yia, Ghorbani:2017jls, Chiang:2017nmu, Ghorbani:2019itr} or without scalar mixing~\cite{Espinosa:1993bs, Espinosa:2011ax, Kakizaki:2015wua, Hashino:2016rvx} etc. has been studied. In the model of electroweak baryogenesis, the condition of strongly first-order phase transition (the sphaleron decoupling condition) can be approximately described by~\cite{Kuzmin:1985mm}
\begin{align}
  \label{eq:appro_sphcondition}
  \frac{v_{c}}{T_{c}} > 1,
\end{align}
where $T_{c}$ is the critical temperature, and $v_{c}$ is the value of the order parameter at $T_{c}$. 

Phenomenological consequences of extended Higgs sectors have been examined at current and future collider experiments. For instance, the two Higgs doublet model (THDM)~\cite{Kanemura:2004mg, Arhrib:2003rp, Asakawa:2010xj, Kanemura:2014dea, Kanemura:2014dja, Kanemura:2014bqa, Bernon:2015qea, Bernon:2015wef, Krause:2016oke, Chowdhury:2017aav, Braathen:2019pxr, Braathen:2019zoh, Aiko:2020ksl}, the SM with singlet fields~\cite{Lopez-Val:2014jva, Robens:2015gla, Bojarski:2015kra, Kanemura:2015fra, Kanemura:2016lkz, Robens:2016xkb, Lewis:2017dme, Adhikari:2020vqo}, the triplet Higgs model~\cite{Aoki:2012jj, Arbabifar:2012bd, Kanemura:2013mc, BhupalDev:2013xol, Blunier:2016peh, Chabab:2018ert} and the inert doublet model (IDM)~\cite{Aoki:2013lhm, Arhrib:2015hoa, Kanemura:2016sos, Kalinowski:2018ylg} have been studied. 
In particular, the triple Higgs boson coupling in several extended Higgs models, which characterize the structure of the Higgs potential, can deviate significantly from the SM prediction via quantum corrections~\cite{Kanemura:2004mg, Kanemura:2016lkz, Braathen:2019pxr, Braathen:2019zoh, Aoki:2012jj, Arhrib:2015hoa}. The large quantum effect is often called \textit{the non-decoupling effect}. New effective field theories describing the non-decoupling effects have recently been proposed~\cite{Falkowski:2019tft, Cohen:2020xca, Kanemura:2021fvp, Kanemura:2022txx}. 
Such large deviations in the triple Higgs boson coupling due to the quantum non-decoupling effects are often predicted in models of electroweak baryogenesis to satisfy the sphaleron decoupling condition~\eqref{eq:appro_sphcondition}~\cite{Grojean:2004xa, Kanemura:2004ch}. 
Namely, the strongly first-order electroweak phase transition can be tested by detecting a large deviation in the triple Higgs boson coupling from the SM prediction at future hadron colliders and lepton colliders such as the High Luminosity-LHC (HL-LHC)~\cite{Liss:2013hbb, CMS:2013xfa}, Future Circular Collider~(FCC-hh)~\cite{FCC:2018vvp}, International Linear Collider~(ILC)~\cite{LCCPhysicsWorkingGroup:2019fvj} and Compact LInear Collider~(CLIC)~\cite{CLIC:2016zwp}.
For example, it has been shown that the electroweak baryogenesis can be realized in the framework of a CP-violating THDM without the constraints from the electric dipole moments~\cite{Kanemura:2020ibp, Enomoto:2021dkl}. In this model, the triple Higgs boson coupling should deviate from the SM prediction about 33-55 \%. 

It has also been known that gravitational waves (GWs) from the first-order phase transition can be used to explore such a scenario of electroweak baryogenesis~\cite{Grojean:2006bp}\footnote{It has been discussed that the observation of primordial black holes may be important as a new tool to verify the first-order electroweak phase transition via cosmological observations \cite{Hashino:2021qoq}. 
}. 
The spectrum takes a special shape with a peak around $10^{-3}$ to $10^{-1}$~Hz. Such GWs are expected to be observed at the LISA~\cite{LISA:2017pwj}, DECIGO~\cite{Seto:2001qf}, BBO~\cite{Corbin:2005ny}, TianQin~\cite{TianQin:2015yph} and Taiji~\cite{Ruan:2018tsw}. From detailed measurements of the GWs, not only the nature of electroweak phase transition~\cite{Grojean:2006bp} but also the structure of the extended Higgs sector may be able to be determined~\cite{Kakizaki:2015wua, Hashino:2016rvx, Ahriche:2018rao, Hashino:2018wee, Gould:2019qek}. 

It has been known that from the unitarity argument \cite{Lee:1977eg} there are upper bounds on the masses of the additional Higgs bosons if the coupling constants of the lightest SM-like Higgs boson $h$ deviate from the SM ones \cite{Kanemura:2014bqa, Kanemura:2015ska, Kanemura:2019kjg}. In the alignment limit where the lightest Higgs boson behaves exactly like the SM Higgs boson at the tree level, on the contrary, no such upper bound is obtained, and the masses of additional Higgs bosons can be very large. 

In this letter, the first-order electroweak phase transition is discussed in models with extended Higgs sectors for the case with relatively heavy additional Higgs bosons. We here employ the more exact expression of the sphaleron decoupling condition. We then examine the parameter space of the THDM where the sphaleron decoupling condition is satisfied with perturbative unitarity \cite{Kanemura:1993hm, Akeroyd:2000wc, Ginzburg:2005dt} and vacuum stability \cite{Deshpande:1977rw}.
A similar analyses are also performed in the model with $N$ singlet scalar fields possessing a $O(N)$ global symmetry ($N$-scalar singlet model)~\cite{Kakizaki:2015wua} and the IDM~\cite{Barbieri:2006dq}. We find that mass upper bounds on additional Higgs bosons are obtained to be at the TeV scale even in the alignment limit. 

We also discuss phenomenological impacts of the case with the additional Higgs bosons with the mass near 1 TeV in these extended Higgs models. We find that even though they are too heavy to be directly detected at current and future experiments at hadron colliders, the large deviation in the triple Higgs boson coupling can be a signature for the first-order electroweak phase transition due to quantum effects of such heavy additional Higgs bosons,  while GWs from the first-order phase transition is weaker in this case as compared to the case with light additional Higgs bosons. 

The structure of this paper is as follows. In Section 2, we give a review of the THDM. In Section 3, we discuss the condition of the strongly first-order electroweak phase transition. Then, in Section 4, we discuss the constraint on the THDM by utilizing the condition defined in Section 3. In Section 5, we consider three benchmark points to show phenomenological differences between the models with light and heavy additional Higgs bosons. We also discuss the triple Higgs boson coupling with relatively heavy additional Higgs bosons in the THDM, and show the GW spectra in each benchmark point. In Section 6, we show a similar discussion on the $N$-scalar singlet model and the IDM. Discussions and conclusions are given in Section 7.

\section{The Two Higgs doublet model \label{sec:THDM}}

We here define the THDM, by using which we explain details of our analysis for the phase transition. For the results in the other models such as the $N$-scalar singlet model and the IDM, we only summarize them in Sec.~\ref{sec:IDM_HSM}. 

We consider the CP-conserving THDM with a softly-broken $Z_{2}$ symmetry $\Phi_{1} \to \Phi_{1}$, 
$\Phi_{2} \to - \Phi_{2}$. The symmetry plays a role to avoid flavor changing neutral currents at the tree level~\cite{Glashow:1976nt}. The Higgs potential in the model is given by
\begin{align}
  \label{eq:potential_THDM_tree}
  \begin{aligned}
     V_{\rm tree}^{\rm THDM}(\Phi_{1},\Phi_{2}) =
     &~  m_{1}^{2}\left|\Phi_{1}\right|^{2}+m_{2}^{2}\left|\Phi_{2}\right|^{2}
   	-\left(m_{12}^{2} \Phi_{1}^{\dagger} \Phi_{2}+\mathrm{h.c.}\right)
	+\frac{\lambda_{1}}{2}\left|\Phi_{1}\right|^{4}
   	+\frac{\lambda_{2}}{2}\left|\Phi_{2}\right|^{4} \\
     &+\lambda_{3} \left|\Phi_{1}\right|^{2}\left|\Phi_{2}\right|^{2} 
       +\lambda_{4}\left|\Phi_{1}^{\dagger} \Phi_{2}\right|^{2}
   	+\left[\frac{\lambda_{5}}{2}\left(\Phi_{1}^{\dagger} \Phi_{2}\right)^{2}
   	+\mathrm{h.c.}\right].
  \end{aligned}
\end{align}
Although $m_{12}^2$ and $\lambda_{5}$ are complex in general, we here assume that these are real for simplicity. 
The doublets $\Phi_{i}~(i=1,2)$ are parameterized as
\begin{align}
  \Phi_{i}=\left(\begin{array}{c}
   			w^{+}_{i} \\
			\frac{1}{\sqrt{2}}(v_{i}+h_{i} + i z_{i})
		\end{array}\right) \quad (i =1,2),
\end{align}
where $\tan \beta = v_{2}/v_{1}$, $v_{1} = v \cos \beta$, $v_{2} = v \sin \beta $ and $v = \sqrt{v_{1}^2 + v_{2}^2}~( \simeq 246 \text{GeV})$.
We reduce two parameters $m_{1}^2$ and $m_{2}^2$ by using the stationary conditions
\begin{align}
  \left. \frac{\partial V_{\rm tree}^{\rm THDM}}{\partial h_{1}} \right|_{\rm min} = 
  \left. \frac{\partial V_{\rm tree}^{\rm THDM}}{\partial h_{2}} \right|_{\rm min} = 0.
\end{align}

\begin{table}
\begin{tabular}{|c||c|c|c|c|c|c|c|}
\hline & $\Phi_{1}$ & $\Phi_{2}$ & $Q_{L}$ & $L_{L}$ & $u_{R}$ & $d_{R}$ & $e_{R}$ \\
\hline \hline \text { Type-I } & + & $-$ & + & + & $-$ & $-$ & $-$ \\
\hline \text { Type-II } & + & $-$ & + & + & $-$ & + & + \\
\hline \text { Type-X } & + & $-$ & + & + & $-$ & $-$ & + \\
\hline \text { Type-Y } & + & $-$ & + & + & $-$ & + & $-$ \\
\hline
\end{tabular}
\caption{
$Z_{2}$ charge assignment in each type of the THDM. 
\label{table:Z2_charge}}
\end{table}
Diagonalizing the mass matrices and introducing the mixing angle $\alpha$ for the CP even neutral scalars,  
we obtain five mass eigenstates; two CP even states ($h, H$), a CP odd state ($A$) and charged states ($H^{\pm}$).
We take $h$ as the Higgs boson discovered at the LHC. The free parameters are given by\footnote{We utilize the definition described in Ref. \cite{Kanemura:2004mg}.}
\begin{align}
  m_{H},~ m_{A},~ m_{H^{\pm}},~ \tan \beta,~ M^2 \equiv m_{12}^2/(\sin \beta \cos \beta), ~ \sin (\beta - \alpha). 
\end{align}
The THDM is classified by the $Z_{2}$ charge assignment for the quarks and charged leptons as shown in Tab.~\ref{table:Z2_charge}. 
We especially focus on the Type-I and Type-II THDM in this paper.

We consider the bound from perturbative unitarity~\cite{Lee:1977eg} to discuss the constraints on the THDM~\cite{Kanemura:1993hm, Akeroyd:2000wc, Ginzburg:2005dt}.
The dimensionless parameters $\lambda_{i} ~ (i=1, \cdots, 5)$ in the Higgs potential in Eq.~\eqref{eq:potential_THDM_tree} are constrained by perturbative unitarity. 
Unless the Higgs field $h$ behaves like the SM Higgs boson, upper bounds on the masses of the additional Higgs bosons are obtained by perturbative unitarity~\cite{Kanemura:2014bqa, Kanemura:2015ska, Kanemura:2015fra, Kanemura:2019kjg}. On the contrary, if $h$ is SM-like, no upper bound on the masses of the additional Higgs bosons is obtained. As we discuss later, the upper bound can be obtained even in such cases by imposing the sphaleron decoupling condition in addition to the unitarity bound. Another theoretical constraint comes from vacuum stability, which is expressed by~\cite{Nie:1998yn, Kanemura:1999xf}
\begin{align}
  \lambda_{1}>0,~ \lambda_{2}>0,~ \lambda_{3}+\lambda_{4}+|\lambda_{5}| > - \sqrt{\lambda_{1} \lambda_{2}},~
  \lambda_{3}>-\sqrt{\lambda_{1} \lambda_{2}}.
\end{align}

The direct searches at collider experiments also set the bound on the masses of the additional Higgs bosons. By the LEP experiments~\cite{ALEPH:2013htx}, the THDM with $m_{H^{\pm}}<78~\text{GeV}$ is ruled out. The additional Higgs bosons are also explored by the LHC experiments. The lower bounds on the masses of the additional Higgs bosons are determined via the $A \to \tau \overline{\tau}$ and $A \to t \overline{t}$ processes \cite{Aiko:2020ksl}. For instance, in the Type-I THDM with $\tan \beta =1$, the mass regions $m_{\Phi}<600$ GeV are excluded where $\Phi = H, A, H^{\pm}$. 
In the Type-II THDM with $\tan \beta < 2 ~ (\tan \beta > 10)$, the mass regions $m_{\Phi}<350$ GeV ($m_{\Phi}<400$ GeV) are excluded.

The masses of the charged Higgs bosons are strongly constrained by flavor experiments~\cite{Haller:2018nnx}. 
In the Type-I THDM with $\tan \beta < 1.5$, $m_{H^{\pm}}<300$ GeV is excluded via the measurement of the $B_{s} \to \mu \mu$ process. For the Type-II THDM, $m_{H^{\pm}}<590$ GeV is excluded independently of $\tan \beta$ via the measurement of the $B \to X_{s} \gamma$ process. 

The measurement of the Higgs boson couplings at the LHC is also important. For the Type-I THDM with $\tan \beta = 1~(\tan \beta =2)$,  $|\cos (\beta - \alpha)| > 0.18 ~ (0.25)$ is excluded. For the type-II THDM, $|\cos (\beta - \alpha)| > 0.09 ~ (0.11)$ is excluded at $\tan \beta = 1~(\tan \beta =2)$. 

Another parameter that is important when discussing constraints on the Higgs sector is the oblique parameters $S$, $T$ and $U$ \cite{Peskin:1991sw}. The experimental constraints on these parameters are given by \cite{Haller:2018nnx}
 \begin{align}
   S = 0.04 \pm 0.11, ~ T = 0.09 \pm 0.14, ~ U = -0.02 \pm 0.11.
 \end{align} 
On the other hand, the two-point function of $W$ and $Z$ bosons in the THDM are calculated in Refs.~\cite{Toussaint:1978zm, Bertolini:1985ia, Grimus:2008nb}. The theoretical calculations and the measurements of the rho parameter indicate that the following condition should be satisfied approximately
\begin{align}
  m_{H^{\pm}} \simeq m_{A} ~~ \text{or} ~~ m_{H^{\pm}} \simeq m_{H} ~\text{with}~ \sin (\beta - \alpha)=1.
\end{align}
This condition is satisfied when the Higgs potential possesses a custodial symmetry \cite{Pomarol:1993mu, Gerard:2007kn, deVisscher:2009zb}.

\section{Condition of Strongly first-order phase transition \label{sec:SFOPT}}

In this section, we discuss the sphaleron decoupling condition. In order to formulate the condition, we should consider the effective potential at finite temperatures. We follow the definition for the effective potential in the THDM in the Parwani scheme~\cite{Parwani:1991gq} discussed in Ref.~\cite{Basler:2016obg}. We also utilize the definition of the nucleation temperature $T_{n}$ described in Ref.~\cite{Grojean:2006bp}. 
We use the public code \texttt{CosmoTransitions} to obtain $T_{n}$ for our numerical evaluation~\cite{Wainwright:2011kj}.

The key of electroweak baryogenesis is a sphaleron transition process. This process violates the baryon number via the chiral anomaly \cite{Klinkhamer:1984di}. To generate the observed baryon asymmetry via the mechanism of electroweak baryogenesis, the sphaleron process must decouple in the broken phase. The transition rate of the sphaleron process is related to the energy of sphalerons at finite temperatures. In order to discuss the feasibility of electroweak baryogenesis, we should evaluate the sphaleron energy in extended Higgs models. 

The sphaleron is a non-perturbative solution in field equations of the $SU(2)$ gauge theory~\cite{Manton:1983nd, Klinkhamer:1984di, Akiba:1988ay}. The sphaleron in extended Higgs models has been calculated~\cite{Moreno:1996zm, Ahriche:2007jp, Funakubo:2009eg, Fuyuto:2014yia, Fuyuto:2015jha, Spannowsky:2016ile, Gan:2017mcv, Chiang:2017nmu, Zhou:2020xqi, Kanemura:2020yyr}. We propose a new ansatz for the configuration of the sphaleron, which is an extension of the ansatz proposed by Spannowsky and Tamarit~\cite{Spannowsky:2016ile} to the finite temperature systems 
\begin{align}
    W_{i}^{a}(\vec{\xi}) &= \frac{v(T)}{2}\left[\epsilon^{a i j} n_{j}\frac{1-R(\xi) \cos \theta(\xi)}{\xi} +\left(\delta_{a i}-n_{a} n_{i}\right) 
    \frac{R(\xi) \sin \theta(\xi)}{\xi}\right],\\
    \Phi_{1}(\vec{\xi})&=\frac{v_{1}(T)}{\sqrt{2}} S_{1}(\xi) e^{i n_{a} \sigma^{a} \phi_{1}(\xi)}\left(\begin{array}{l}0 \\ 1\end{array}\right), \\
     \Phi_{2}(\vec{\xi})&=\frac{v_{2}(T)}{\sqrt{2}} S_{2}(\xi) e^{i n_{a} \sigma^{a} \phi_{2}(\xi)}\left(\begin{array}{l}0 \\ 1\end{array}\right),
\end{align}
where $\vec{\xi} = g v(T) \vec{r}/2$, $\xi = |\vec{\xi}|$ and $ v(T) = \sqrt{v_{1}(T)^2 + v_{2}(T)^2}$. The profile functions $R$, $S_{1}$ and $S_{2}$ satisfy the following boundary conditions
\begin{align}
  \begin{aligned}
     \lim_{\xi \to 0}R(\xi) \to -1, ~  \lim_{\xi \to 0}S_{1}(\xi) \to 0, ~  \lim_{\xi \to 0} S_{2}(\xi) \to 0, \\
     \lim_{\xi \to \infty}R(\xi) \to 1, ~  \lim_{\xi \to \infty}S_{1}(\xi) \to 1, ~  \lim_{\xi \to \infty} S_{2}(\xi) \to 1.
  \end{aligned}
  \label{eq:sph_BC}
\end{align}
We take $\theta(\xi) = \pi$ and $\phi_{i}(\xi) = \pi/2 ~(i=1,2)$ as taken in Ref.~\cite{Spannowsky:2016ile}. 
The sphaleron energy at finite temperatures $E_{\rm sph}(T)$ is given by
\begin{align}
  E_{\rm sph}(T) &= \frac{4 \pi v(T)}{g} \mathcal{E}(T), \\
   \mathcal{E}(T) &= \int d \xi \left[ \frac{1}{2}\left( \frac{\partial R}{ \partial \xi} \right)^2 
   + \frac{1}{4 \xi^2}\left( 1-R(\xi) \right)^2 
   + \frac{v_{1}(T)^2 \xi^2}{v(T)^2} \left\{ \left( \frac{\partial S_{1}}{\partial \xi} \right)^2 
   + \frac{1}{2 \xi^2} S_{1}^2 (1-R(\xi))^2 \right\}  \right. \nonumber \\ &\left.
   + \frac{v_{2}(T)^2 \xi^2}{v(T)^2} \left\{ \left( \frac{\partial S_{2}}{\partial \xi} \right)^2 
   + \frac{1}{2 \xi^2} S_{2}^2 (1-R(\xi))^2 \right\} 
   + \frac{8 \xi^2}{ g^2 v(T)^4} \left( V_{\rm eff}(S_{1},S_{2}, T) - V_{\rm eff}(v_{1},v_{2}, T) \right) \right], 
\end{align}
where $V_{\rm eff}$ is the effective potential at finite temperatures. The profile functions $R,~S_{1}$ and $S_{2}$ are determined to realize the saddle point of the energy functional and satisfy the boundary conditions described in Eq.~\eqref{eq:sph_BC}.

The condition for the suppression of the baryon number violating process in the broken phase is given by
\begin{align}
  \label{eq:sphaleron_decoupling1}
  - \frac{1}{N_B} \frac{dN_B}{dt}
  \simeq A(T) e^{-E_{\rm sph}(T)/T} < H_{\rm Hubble}(T).
\end{align}
where $N_{B}$ is the baryon number, and $H_{\rm Hubble}(T)$ is the Hubble parameter at $T$. The prefactor $A(T)$ is a fluctuation determinant defined around the sphaleron configuration~\cite{Arnold:1987mh}. Using $A(T)$ calculated within the SM, the inequality~\eqref{eq:sphaleron_decoupling1} evaluated at $T=T_{n}$ is transformed into~\cite{Funakubo:2009eg, Gan:2017mcv}
\begin{align}
  \frac{v(T_{n})}{T_{n}} 
  > \frac{g}{4 \pi \mathcal{E}(T_{n})} \left[ 41.65 + 7 \ln \frac{v(T_{n})}{T_{n}} - \frac{T_{n}}{100{\rm GeV}} \right]
  \equiv \zeta_{\rm sph}(T_{n}).
  \label{eq:sph_decoupling_condition}
\end{align}
We take the above condition as the criterion for the strongly first-order phase transition. In the following, we discuss the constraint on the several extended Higgs models by utilizing the condition.

We comment on the thermal correction to the effective potential by new particles. For the THDM with the alignment, field dependent masses of additional Higgs bosons are given by $m_{\Phi}^2(\phi) = M^2 + \lambda_{\Phi} \phi^2~(\Phi = H, A, H^{\pm})$. $\phi$ is the order parameter, and $\lambda_{\Phi}$ is the linear combination of $\lambda_{i}~(i=1,\cdots,5)$ in Eq.~\eqref{eq:potential_THDM_tree}. On the other hand, the thermal correction to the effective potential has the Boltzmann suppression factor $\exp \left[ -m_{\Phi}^2(\phi)/T^2 \right]$~\cite{Basler:2016obg}. In the decoupling region ($M^2 \gg \lambda_{\Phi} v^2$), the Boltzmann suppression is significant in the thermal correction. On the contrary, in the non-decoupling region ($M^2 \lesssim \lambda_{\Phi} v^2$), the Boltzmann suppression factor is $O(1)$ at $\phi=0$.

\section{Bounds on masses of the additional Higgs bosons \label{sec:Mmax_THDM}}

\begin{figure}[t]
  \begin{center}
    \includegraphics[width=16cm]{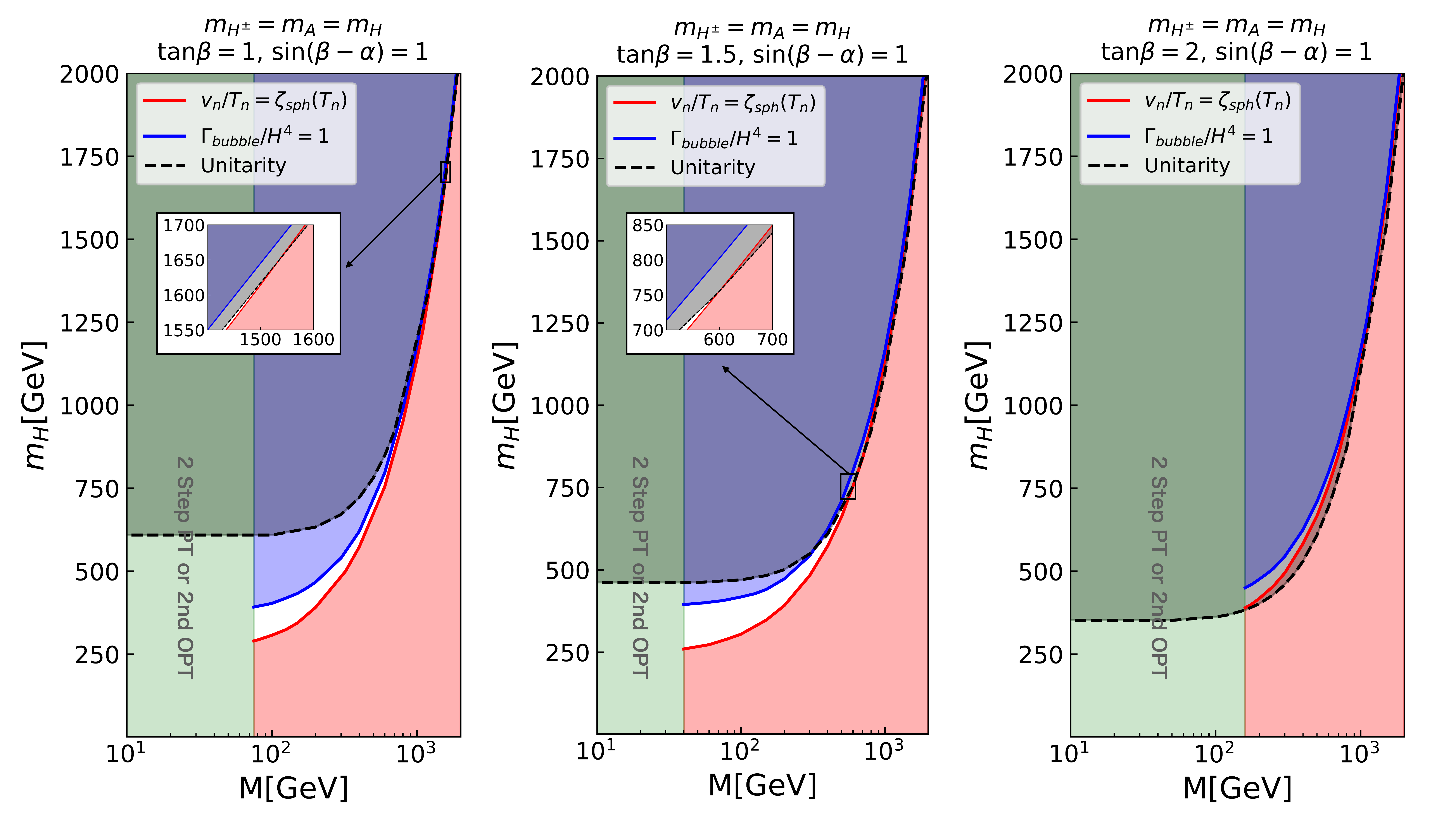}
     \caption{
     The allowed parameter regions for $m_{H^{\pm}} = m_{A} = m_{H}, ~ \sin (\beta - \alpha)=1$ and
     $\tan \beta =1,~ 1.5 ~ \text{and} ~ 2$.
     The red region is excluded by the sphaleron decoupling condition. The blue region represents that the
     electroweak phase transition has not been completed. The gray region is excluded by the unitarity bound.
     The green region indicates that the phase transition is two step where the first phase transition is a
     second-order phase transition, or the single step second-order phase transition. 
     The strongly first-order electroweak phase transition cannot be realized in Scenario~1 with $\tan \beta > 2$.
     \label{fig1}
     }
  \end{center}
\end{figure}

In this section, we discuss the constraint on the THDM by using the sphaleron decoupling condition \eqref{eq:sph_decoupling_condition} in the following several cases.

\noindent
{\bf [Scenario 1]} (Alignment with degenerated masses)

We here discuss the scenario in which all the coupling constants of the Higgs boson $h$ are SM-like, and the masses of the additional Higgs bosons are degenerate. In Fig.~\ref{fig1}, parameter regions are shown where the sphaleron decoupling condition in Eq.~\eqref{eq:sph_decoupling_condition} is satisfied. 
In the red region, the sphaleron decoupling condition is not satisfied. 
For the heavy mass region $m_{\Phi} > 1$TeV $(\Phi = H, A, H^{\pm})$, the Boltzmann suppression in the thermal correction is significant because $M$ is large. In such a case, the strongly first-order phase transition can be realized mainly by the radiative correction to the effective potential at the zero temperature~\cite{Dorsch:2017nza}.
The blue region is excluded by the condition for the completion of electroweak phase transition.\footnote{This condition is given by $\Gamma_{\rm bubble}(T)/H_{\rm Hubble}(T)^4 = 1$ at $T = T_{n}$ where $\Gamma_{\rm bubble}(T)$ is the nucleation rate of the vacuum bubbles. } 
In the gray region, the unitarity bound is not satisfied. The green region indicates that the phase transition is two step where the first phase transition is a second-order phase transition, or single step second-order phase transition.
The upper bounds on the additional Higgs boson masses are determined by the combination of the sphaleron decoupling condition and the unitarity bound. 

In the alignment limit, there is no upper bound on the masses of the additional Higgs bosons without imposing the sphaleron decoupling condition \cite{Kanemura:2014bqa, Kanemura:2015ska, Kanemura:2015fra, Kanemura:2019kjg}. Even in such a case, the upper bound is obtained by combining the sphaleron decoupling condition with the unitarity bound.
\begin{figure}[t]
  \begin{center}
    \includegraphics[width=16cm]{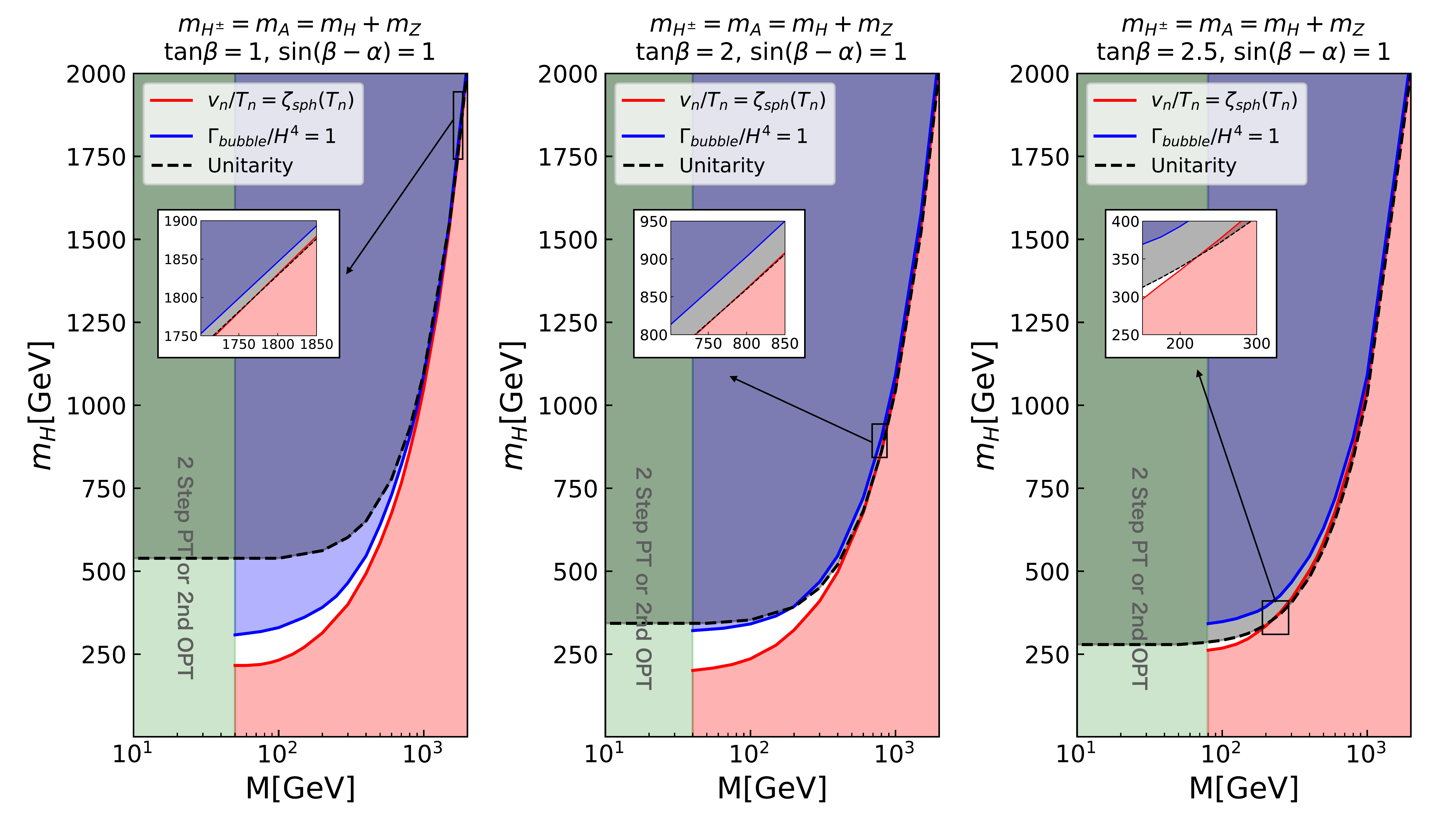}
     \caption{
     The allowed regions in the THDM with $m_{H^{\pm}} = m_{A} = m_{H} + m_{Z}, ~ \sin (\beta - \alpha)=1$ and 
     $\tan \beta =1,~ 2,~ 2.5$. The definition of the regions for each color is the same in Fig. \ref{fig1}.
     \label{fig2}
     }
  \end{center}
\end{figure}\\

\noindent
{\bf [Scenario 2]} (Alignment with a relatively small mass difference)

Next, we consider the THDM with $m_{H^{\pm}} = m_{A} = m_{H} + m_{Z}$ and $\sin (\beta - \alpha)=1$. In Fig.~\ref{fig2}, the allowed parameter regions in Scenario 2 are shown. These regions satisfy the same conditions as in Fig. \ref{fig1}. If the masses of the additional Higgs bosons are not degenerate, we cannot take the limit $m_{\Phi} \to M$, where $\Phi = H,~ A,~ H^{\pm}$. Hence, it is not possible to take large $m_{\Phi}$ while keeping the dimensionless parameters $\lambda_{i} ~ (i=1, \cdots, 5)$ small. Therefore, in such a case, the upper bounds on the masses of the additional Higgs bosons are determined only by the argument of perturbative unitarity. However, we have confirmed that the bound is weaker than the bound obtained by the combination of the sphaleron decoupling condition and perturbative unitarity in Scenario~2. 
\\

\noindent
{\bf [Scenario 3]} (Alignment with a relatively large mass difference)

\begin{figure}[t]
  \begin{center}
    \includegraphics[width=8cm]{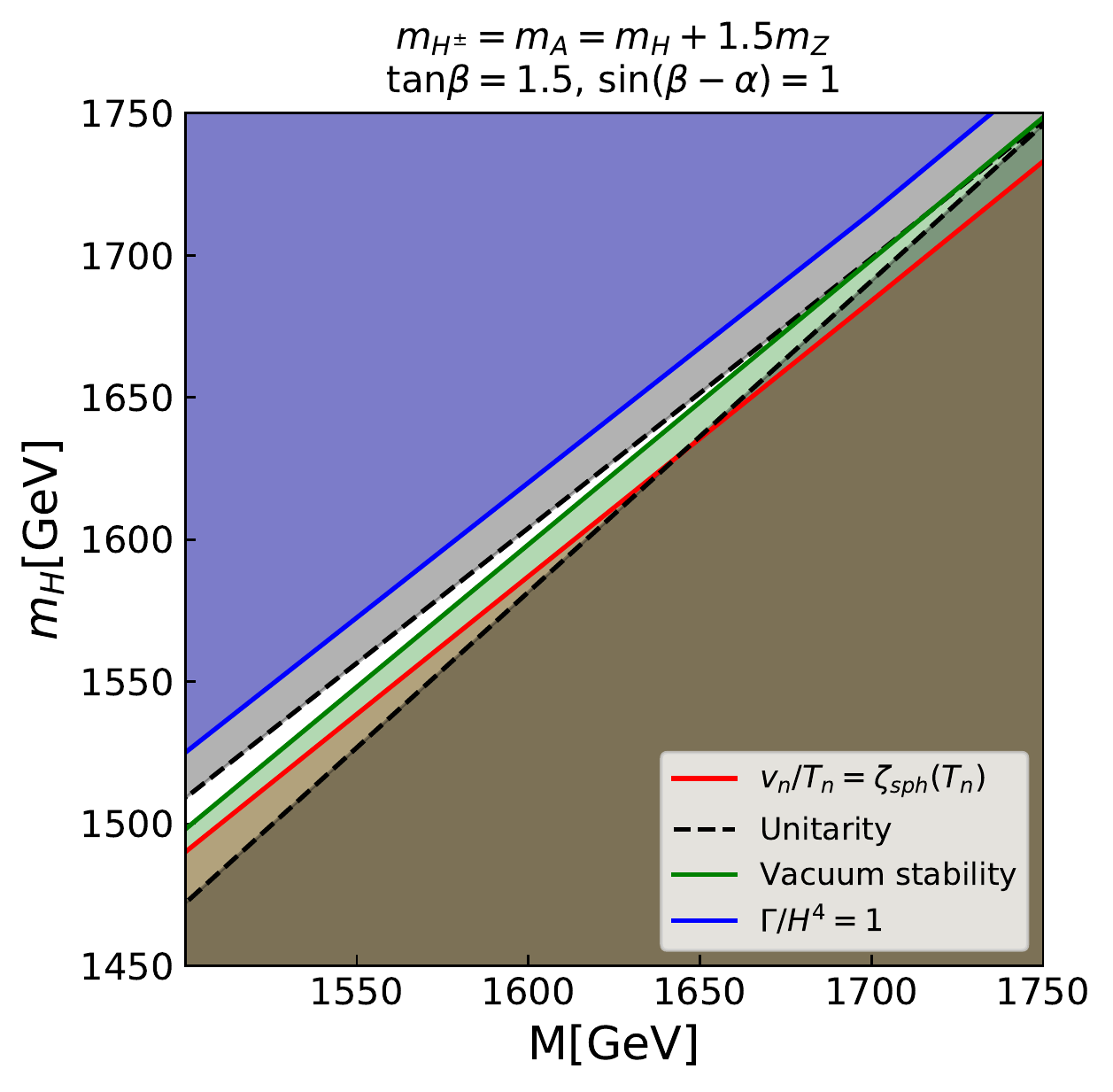}
     \caption{
     The allowed parameter regions in THDM with 
     $m_{H^{\pm}} = m_{A} = m_{H} + 1.5 m_{Z}, ~ \sin (\beta - \alpha)=1$ and $ \tan \beta = 1.5$.
     In addition to the theoretical constraints described in Fig.~\ref{fig1}, the constraint
     from the vacuum stability is shown. 
     Upper bounds on the additional Higgs boson masses are given by perturbative unitarity and vacuum stability in Scenario~3.
     \label{fig3}
     }
  \end{center}
\end{figure}

We consider the THDM with $m_{H^{\pm}} = m_{A} =  m_{H} + 1.5 m_{Z}, ~ \sin (\beta - \alpha)=1$ and $ \tan \beta = 1.5$.
In Fig. \ref{fig3}, we show the allowed parameter regions in Scenario 3 at large $M$ regions. In this case, the upper bounds on the masses of the additional Higgs bosons cannot be determined by the sphaleron decoupling condition and the unitarity bound. Instead, the upper bounds are determined by perturbative unitarity and vacuum stability. As we discussed in the case of Scenario 2, when the additional Higgs bosons have a large mass difference, we cannot take large $m_{\Phi}$ . The theoretical constraints on the masses of the additional Higgs bosons are stronger as the mass difference increases. 
\\

\noindent
{\bf [Scenario 4]} (Non-alignment)

\begin{figure}[t]
  \begin{center}
    \includegraphics[width=10cm]{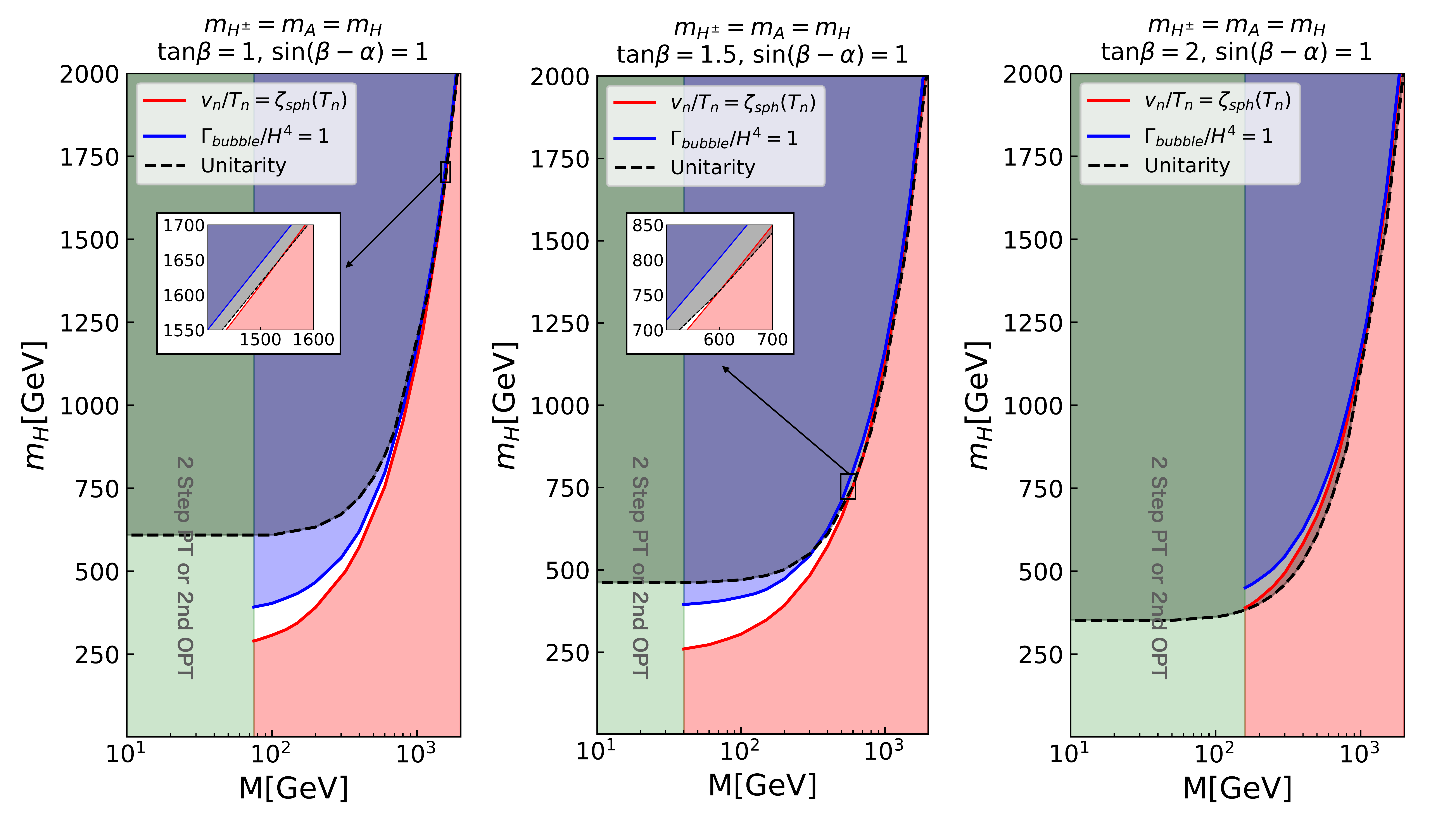}
     \caption{
     The allowed parameter regions in the THDM for $\sin (\beta - \alpha)=1$ and $\sin (\beta - \alpha)=0.999$.
     For comparison, we show the left figure in Fig.~\ref{fig1} again as the left panel.
     The upper bounds on the masses of the additional Higgs bosons are around 1.2 TeV for $\sin (\beta - \alpha)=0.999$.
     The definition of the regions for each color is the same in Fig.~\ref{fig1}.
     \label{fig4}
     }
  \end{center}
\end{figure}

Finally, we discuss the THDM without alignment. As an example, we focus on the model with $m_{H^{\pm}}=m_{A}=m_{H}, ~ \tan \beta = 1 $ and $ \sin (\beta - \alpha)=0.999$. In Fig.~\ref{fig4}, we can see the importance of the alignment.  In this case, the constraint on the masses of the additional Higgs bosons is more stringent when $\sin (\beta - \alpha)$ deviates from unity. Thus, the mass upper bounds are lower than those in the case with alignment. This result indicates that the strongly first-order phase transition in the THDM with relatively heavy additional Higgs bosons prefers the alignment. 

As shown in this section, even when we consider the THDM with the alignment, we can obtain the upper bounds on the masses of the additional Higgs bosons by using the sphaleron decoupling condition.
We have numerically confirmed that the upper bounds are around 1.6-2 TeV. 
If no new scalar particles are discovered below this bound, realizing the scenario of electroweak baryogenesis may be difficult. Our result provides an important criterion for verifying the feasibility of electroweak baryogenesis.

Before closing this section, we give two comments on our analysis.
The strongly first-order electroweak phase transition in the THDM with the heavy additional Higgs bosons requires relatively large $\lambda_{i} ~ (i=1, \cdots, 5)$. In such a case, the sub-leading terms neglected in our thermal mass calculation may be non-negligible due to additional contributions from super daisy diagrams \cite{Curtin:2016urg}. Since there is no established method to systematically incorporate these effects in the THDM, we have only taken into account thermal masses, as often done so in the literature~\footnote{
For several extended Higgs models, systematic methods are discussed to include the sub-leading finite temperature corrections to the thermal mass~\cite{Croon:2020cgk, Schicho:2021gca, Niemi:2021qvp}.
}.

We have obtained the upper bounds on the masses of the additional Higgs bosons. Getting this bound, we have used the unitarity bound at the tree level and the sphaleron decoupling coupling condition using the effective potential at the one-loop level. When higher-order corrections are considered, upper bounds on masses of additional Higgs bosons can be changed. However, it is expected that even in such case upper bounds on masses of additional Higgs bosons can exist.

\section{Signatures for relatively heavy additional Higgs bosons \label{sec:GW_hhh}}

\begin{table}[t]
\begin{tabular}{|c|c|c|c|c|c|c|c|c|}
\hline
 & $m_{H^{\pm}}$ & $m_{A}$ & $m_{H}$ & $M$ & $\tan \beta$ & $\Delta \lambda_{hhh}^{1 \ell}/\lambda_{hhh}^{\rm SM}$ & $\Delta \lambda_{hhh}^{2 \ell}/\lambda_{hhh}^{\rm SM} $ & $v_{n}/T_{n}$ \\ \hline
BM0 & 373GeV & 373GeV & 373GeV & 50GeV & 1 & 71.5\% & 86.4\% & 3.80 \\ \hline
BM1 & 464GeV & 464GeV & 373GeV & 200GeV & 1.8 & 80.2\% & 112\% & 2.60 \\ \hline
BM2 & 891GeV & 891GeV & 800GeV & 720GeV & 1.8 & 80.2\% & 125\% & 2.37 \\ \hline
\end{tabular}
\caption{ Benchmark scenarios with $\sin (\beta - \alpha) =1$. $\Delta \lambda_{hhh}^{1 \ell}/\lambda_{hhh}^{\rm SM}$ is the deviation in the triple Higgs boson coupling at the one-loop level. $\Delta \lambda_{hhh}^{2 \ell}/\lambda_{hhh}^{\rm SM}$ is that at the two-loop level. }
\label{table:benchmark}
\end{table}

As we have shown in Sec.~\ref{sec:Mmax_THDM}, in the THDM with the additional Higgs bosons whose masses are larger than 1 TeV, the strongly first-order electroweak phase transition may be realized. It would be difficult to test such models at near future collider experiments such as HL-LHC and the ILC. We here discuss how to verify the scenario with heavy additional Higgs bosons. We focus on the three benchmark scenarios BM0, BM1 and BM2 as shown in Tab.~\ref{table:benchmark}. For BM1 the additional Higgs bosons are relatively light (a few 100 GeV), while for BM2 the additional Higgs bosons are relatively heavy (around 1 TeV). As shown in Ref.~\cite{Aiko:2020ksl}, testing BM2 is difficult at the HL-LHC and the ILC. Although BM0 cannot satisfy the experimental constraints from LHC and current flavor experiments, we dare to show the GW spectrum in this benchmark for comparison. 

As we mentioned in the introduction, the triple Higgs boson coupling is the key to verify the first-order electroweak phase transition. The triple Higgs boson coupling is defined by using the effective potential $V_{\rm eff}$ as
\begin{align}
  \lambda_{hhh} \equiv \left. \frac{\partial^3 V_{\rm eff}(h, T=0)}{\partial h^3} \right|_{\rm min}.
\end{align}
We define the deviation in the triple Higgs boson coupling from the SM prediction as $\Delta \lambda_{hhh}/ \lambda_{hhh}^{\rm SM}\equiv (\lambda_{hhh} - \lambda_{hhh}^{\rm SM})/\lambda_{hhh}^{\rm SM}$, where $\lambda_{hhh}^{\rm SM}$ is the value in the SM.
$\Delta \lambda_{hhh}/ \lambda_{hhh}^{\rm SM}$ can be significant even in extended Higgs models with heavy additional Higgs bosons due to their quantum effects~\cite{Kanemura:2004mg, Braathen:2019pxr, Braathen:2019zoh}. We have confirmed that a large $\Delta \lambda_{hhh}/ \lambda_{hhh}^{\rm SM}$ is required to satisfy the sphaleron decoupling condition in the heavy scenario such as BM2. In order to satisfy the sphaleron decoupling condition in the THDM with $m_{\Phi}>700~\text{GeV}~(\Phi = H, A, H^{\pm})$, $\Delta \lambda_{hhh}/\lambda_{hhh}^{\rm SM} > 60 \%$ is required at the one-loop level.
The results indicate that the THDM with relatively heavy additional Higgs bosons can be tested by the measurement of the triple Higgs boson coupling at future collider experiments.

Two-loop corrections to the triple Higgs boson coupling in the THDM have been calculated in Refs.~\cite{Braathen:2019pxr,Braathen:2019zoh}. Including the scalar two-loop corrections, the deviation in the triple Higgs boson coupling is larger than the one-loop result. We have evaluated the constraint on the triple Higgs boson coupling from the sphaleron decoupling condition including the two-loop corrections. Then, we have obtained that $\Delta \lambda_{hhh}/\lambda_{hhh}^{\rm SM} > 80 \%$ is required to satisfy the sphaleron decoupling condition in the THDM with $m_{\Phi}>700 ~ \text{GeV}~ (\Phi = H, A, H^{\pm})$ at the two-loop level. The lower bound on the triple Higgs boson coupling is larger by including the two-loop corrections. 

On the other hand, we do not take into account the two-loop corrections to the strength of the phase transition. According to Refs.~\cite{Laine:2017hdk,Senaha:2018xek}, the strength of the phase transition is weakened by about 10\% due to the two-loop corrections in the IDM. Since the two-loop corrections to the effective potential at finite temperatures in THDM have not been calculated completely, we only consider the effective potential with the one-loop corrections and daisy resummation.

We also discuss the GWs from the first-order electroweak phase transition.
The spectrum of the GWs from the strongly first-order electroweak phase transition is characterized by $\alpha_{\rm GW}$ and $\tilde{\beta}_{\rm GW}$. These parameters are defined by \cite{Grojean:2006bp} 
\begin{align}
  &\alpha_{\rm GW} \equiv 
  \left. \frac{1}{\rho_{\rm rad}} \left[ -\Delta V_{\rm eff} + T \frac{\partial \Delta V_{\rm eff}}{\partial T} \right] \right|_{T=T_{n}^{}},
  ~ \text{where}~~ \rho_{\rm rad}(T) = \frac{\pi^2}{30}g_{*}T^4, \\
  &\tilde{\beta}_{\rm GW} \equiv
  \frac{\beta_{\rm GW}}{ H_{\rm Hubble}(T) }=
  \left. T \frac{d}{dT} \left( \frac{S_{3}}{T} \right) \right|_{T=T_{n}^{}}.
\end{align}
where $\Delta V_{\rm eff} = V_{\rm eff}(\varphi^{B}_{1}(T), \varphi^{B}_{2}(T),T) - V_{\rm eff}(0, 0, T)$. $\varphi^{B}_{i}~(i=1,2)$ is the bounce solutions for the vacuum bubbles. $S_{3}(T)$ is the free energy of the vacuum bubbles. 

GW spectra $\Omega_{\rm GW}(f)$ from the first-order electroweak phase transition consist of three sources; collisions of the vacuum bubbles ($\Omega_{\varphi}$), compressional waves (sound waves) ($\Omega_{\rm sw}$) and magnetohydrodynamics turbulence ($\Omega_{\rm turb}$) \cite{Caprini:2015zlo}
\footnote{
Recently, the effect of strongly first-order phase trainsition on the fitting functions for the GW spectra has been evaluated~\cite{Ellis:2018mja}. In this paper, we discuss the prediction of the GW spectra by using the fitting functions, which have often been used in previous studies.
};
\begin{align}
  h^2 \Omega_{\rm GW}(f) \simeq h^2 \Omega_{\varphi}(f) + h^2 \Omega_{\rm sw}(f) + h^2 \Omega_{\rm turb}(f). 
\end{align}
In general, the leading contribution is the sound wave source $\Omega_{\rm sw}(f)$~\cite{Kakizaki:2015wua, Hashino:2016rvx}. 

\begin{figure}[t]
  \begin{center}
    \includegraphics[width=10cm]{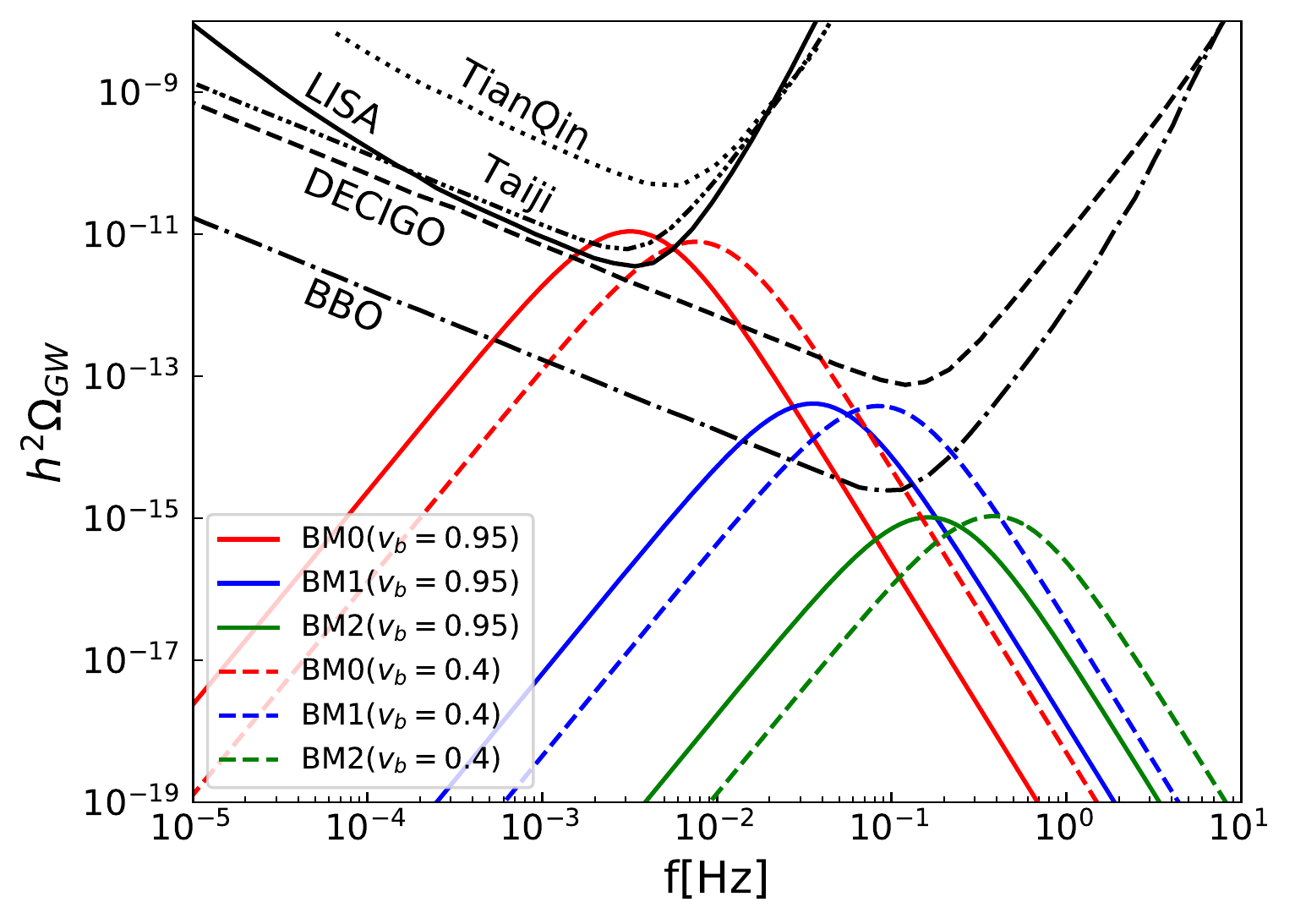}
     \caption{
     The GW spectra in each benchmark scenario.
     The solid (dashed) lines are the cases that the wall velocity is $95\%$ ($40\%$) of the light speed. 
     \label{fig:GWspectrum}
     }
  \end{center}
\end{figure}

We focus on the three benchmarks in Tab.~\ref{table:benchmark}.
In Fig.~\ref{fig:GWspectrum} the GW spectra in each benchmark scenario are shown for the different wall velocity ($v_{b}$). The sensitivity curves at each future GW observation are also shown. The solid (dashed) lines correspond to the cases that the wall velocity is $95\%$ ($40\%$) of the light speed
\footnote{
There are previous studies that have clarified a relation between the wall velocity and the Higgs potential by using quasiclassical calculation methods~\cite{Lewicki:2021pgr}. In this paper, however, the wall velocity is treated as a free parameter.
}.
Interestingly, although magnitudes of the deviation in the triple Higgs boson coupling at the one-loop level are similar between BM1 and BM2, the peak height of the GW spectrum is lower when the additional Higgs bosons are heavy.
If both large $\Delta \lambda_{hhh}/\lambda_{hhh}^{\rm SM}$ and the peaked GW spectrum are determined at future experiments, the additional Higgs bosons are expected to be relatively light. On the other hand, if large $\Delta \lambda_{hhh}/\lambda_{hhh}^{\rm SM}$ is found but no GW spectrum is observed, the scenario with relatively heavy additional Higgs bosons may be plausible.

We note that a detailed analysis for the detectability of the GWs is required in order to determine the mass scale of the additional Higgs bosons. 
We may be able to guess the mass scale of additional Higgs bosons by using the correlation between the GW spectrum and the triple Higgs boson coupling.
The analysis for the detectability of the GW spectrum is beyond the scope of this paper. Although the GW spectrum in BM1 is lower than the sensitivity curves of the LISA, Taiji and DECIGO, we may be able to detect the signal by investigating the sensitivity of these interferometers in details~\cite{Hashino:2018wee}.

\section{Electroweak phase transition in the $N$-scalar singlet model and the Inert doublet model \label{sec:IDM_HSM}}

Following the analysis for the THDM, we here analyze the phase transition in other models such as the $N$-scalar singlet model and in the IDM. 

For simplicity, we consider the model with $N$ additional singlet real scalar fields $S_{i}$ which have a global $O(N)$ symmetry \cite{Espinosa:1993bs}, 
\begin{align}
  V_{N {\rm scalar} }(\Phi, \vec{S})
  = -\mu^2 \Phi^{\dagger} \Phi+ \lambda (\Phi^{\dagger} \Phi)^2 
  + \frac{\mu_{S}^2}{2} |\vec{S}|^2
     + \frac{\lambda_{S}}{4!}|\vec{S}|^4 + \lambda_{\Phi S} |\vec{S}|^2 \Phi^{\dagger} \Phi, 
\end{align}
where $(\vec{S})^{\rm T} = (S_{1}, ..., S_{N})$ is a vector under the $O(N)$ symmetry. We also assume $\mu_{S}^2>0$. In order to obtain upper bounds on the masses of the additional Higgs bosons, we utilize the bound from perturbative unitarity \cite{Cynolter:2004cq} and the sphaleron decoupling condition given in Eq.~\eqref{eq:sph_decoupling_condition}. 
In this model, we obtain the upper bounds on the masses of the additional singlet fields as 2 TeV (1.4 TeV) when $N=1~(N=4)$. As $N$ is larger, this upper bound is more stringent.  

Next, we show the results in the IDM. The Higgs potential is given by
\begin{align}
  V_{\rm IDM}(\Phi_{1},\Phi_{2})
  =& \mu_{1}^2 \Phi_{1}^{\dagger} \Phi_{1} + \mu_{2}^2 \Phi_{2}^{\dagger} \Phi_{2}
  + \frac{\lambda_{1}}{2} (\Phi_{1}^{\dagger} \Phi_{1})^2
  + \frac{\lambda_{2}}{2} (\Phi_{2}^{\dagger} \Phi_{2})^2 \nonumber \\
  &+ \lambda_{3} (\Phi_{1}^{\dagger} \Phi_{1}) (\Phi_{2}^{\dagger} \Phi_{2})
  + \lambda_{4} (\Phi_{1}^{\dagger} \Phi_{2}) (\Phi_{2}^{\dagger} \Phi_{1})
  + \frac{\lambda_{5}}{2} 
   \left[ (\Phi_{1}^{\dagger} \Phi_{2})^2 + {\rm h.c.} \right].
\end{align}
Like the THDM, the IDM has five mass eigenstates; two CP even states~($h, H$), a CP odd state~($A$) and charged states~($H^{\pm}$). To avoid the rho parameter constraint, we take $m_{H^{\pm}} =m_{A}$. We identify the CP even Higgs field $H$ as a dark matter candidate in this paper. By the direct searches such as the LUX \cite{LUX:2013afz}, the dark matter mass ($m_{H}$) is strongly constrained. If we take $m_{H}=m_{h}/2$, we can obtain the constraints on the masses of the charged Higgs bosons ($m_{H^{\pm}}$) and the CP odd Higgs boson ($m_{A}$) from the sphaleron decoupling condition and the completion condition of the phase transition.
The lower bound is determined by the sphaleron decoupling condition. The upper bound is determined by the completion condition of the phase transition; 
\begin{align}
  300 \text{GeV}< m_{H^{\pm}},~ m_{A}< 410 \text{GeV}. 
\end{align}
We note that we have obtained the above lower bound by using the sphaleron decoupling condition given in Eq.~\eqref{eq:sph_decoupling_condition}. It means that our result is the improvement of the previous work~\cite{Gil:2012ya,Borah:2012pu, Fabian:2020hny}.

\section{Discussions and Conclusions \label{discussion_conclusion}}

We give comments on several issues.
We have treated the CP-conserving THDM with softly-broken $Z_{2}$ symmetry. As confirmed in Refs.~\cite{Basler:2021kgq, Fromme:2006cm}, due to the inclusion of non-zero CP-violating phases, strength of the first-order phase transition tends to be weakened. In this case, the constraints on the THDM from the sphaleron decoupling condition might be more stringent than our results. 

We have analyzed the constraint on the extended Higgs models by using the sphaleron decoupling condition, the completion condition of electroweak phase transition, perturbative unitarity and vacuum stability. In addition to these theoretical constraints, if we include the bound from the triviality~\cite{Lindner:1985uk}, the allowed parameter region can be narrowed down in general~\cite{Flores:1982pr, Kominis:1993zc, Nie:1998yn,  Kanemura:1999xf}. Thus, we expect that the upper bounds on the additional Higgs boson masses are lower. However, the mass upper bounds determined by the triviality include a cutoff scale dependence. Therefore, we have not taken into account the triviality as a theoretical constraint. 

We have utilized perturbative unitarity at the tree level to discuss the constraints on the extended Higgs models. When we consider the unitarity bound at the one-loop level, the extended Higgs models might be more strongly constrained~\cite{Grinstein:2015rtl}. But, the unitarity bounds at the one-loop level are inherently energy dependent. In our paper, to obtain the conservative mass upper bounds on the additional Higgs bosons, we have only considered the constraint from perturbative unitarity at the tree level. 

We mention the relation between our results and the predictions in the effective field theories. 
In the Standard Model Effective Field Theory (SMEFT) with a dimension-six operator $|\Phi|^6/\Lambda^2$ where $\Lambda$ is the cutoff scale, the sphaleron decoupling condition requires $\Lambda < 750$GeV as shown in Refs.~\cite{Grojean:2004xa,Delaunay:2007wb}. On the other hand, we have shown that the strongly first-order electroweak phase transitions are possible in the renormalizable extended Higgs models such as the THDM even in the masses of the additional Higgs bosons are above 750 GeV.  It indicates that the strongly first-order electroweak phase transition cannot be comprehensively explored by the SMEFT framework. Instead, the non-linear form of the effective field theory (Higgs EFT) would well describe the strongly first-order phase transition~\cite{Falkowski:2019tft, Cohen:2020xca, Kanemura:2021fvp, Kanemura:2022txx}.

In this paper, in addition to the unitarity bound, we have evaluated the constraint on the extended Higgs models by using the sphaleron decoupling condition given in Eq.~\eqref{eq:sph_decoupling_condition}. In the THDM, we have obtained the new result that the upper bounds on the masses of additional Higgs bosons exist around 1.6-2 TeV even when $h$ is SM-like. 
This indicates that even if the THDM with relatively heavy Higgs bosons whose masses are TeV scale, the strongly first-order electroweak phase transition can be realized. Since light additional Higgs bosons will soon be strongly constrained by future flavor and collider experiments, it might be important to clarify the possibility of the strongly first-order phase transition due to the quantum effects of heavy additional Higgs bosons. 

We have found that in order to realize the strongly first-order phase transition in the THDM with $m_{\Phi}>700 ~ \text{GeV}~(\Phi = H, A, H^{\pm})$, the triple Higgs boson coupling must deviate from the SM prediction at least $80\%$ at the two-loop level. This result is important to verify such scenarios at near future collider experiments such as the HL-LHC and the ILC where the deviation in the triple Higgs boson coupling can be measured.

We have also confirmed that the peak height of the GW spectrum is lower as the masses of the additional Higgs bosons are larger even when the deviation in the triple Higgs boson coupling is similar.
If the large deviation in the triple Higgs boson coupling and the peaked GW spectrum are found, we can expect that the additional Higgs bosons are relatively light. On the contrary, if the large deviation is found in the triple Higgs boson coupling but no GW spectrum is observed, it would be plausible that the additional Higgs bosons are relatively heavy. We may be able to guess the scale of masses of the additional Higgs bosons even if these additional fields are not discovered by direct searches at future collider experiments.

\begin{acknowledgments}

The work of S. K. was supported in part by the Grant-in-Aid on Innovative Areas, the Ministry of Education, Culture, Sports, Science and Technology, No. 16H06492 and JSPS KAKENHI Grant No. 20H00160. 
The work of M. T. was supported in part by JSPS KAKENHI Grant No. JP21J10645.

\end{acknowledgments}

\end{document}